\documentclass{optica-article}

\journal{opticajournal} 

\articletype{Research Article}

\begin{document}

\title{Light focusing through dynamic media via real-valued intensity transmission matrix}

\author{Xuan Liu,\authormark{1} Sebastien Ourselin,\authormark{1} and Tianrui Zhao\authormark{1,*}}

\address{\authormark{1}School of Biomedical Engineering and Imaging Sciences, King’s College London, 4th Floor, Lambeth Wing, St Thomas' Hospital London, London SE1 7EH, United Kingdom}

\email{\authormark{*}tianrui.zhao@kcl.ac.uk} 


\begin{abstract*} 
Precise light delivery through biological tissue is essential for deep-tissue imaging and phototherapeutic applications. Wavefront shaping enables control over scattered light by modulating the incident wavefront, but its application in living tissue is hindered by tissue-induced temporal decorrelation. This study systematically investigated the real-valued intensity transmission matrix (RVITM), a high-speed wavefront shaping method, for light focusing across a broad range of speckle decorrelation times. The inherent trade-off between static light focusing enhancement and implementation speed is characterized, which provided practical guidelines for implementing RVITM in real-time wavefront shaping under varying dynamic conditions. Effective optical focusing using a RVITM with 33 ms runtime was achieved through porcine liver with decorrelation times as short as 13 ms, demonstrating feasibility for biologically relevant dynamics and supporting the development of adaptive, non-invasive optical control for biomedical applications. 

\end{abstract*}

\section{Introduction}
Precise light control in biological tissues is of great importance for a wide range of biomedical applications such as deep-tissue optical contrast imaging, optogenetic stimulation, microsurgery, and photodynamic therapy\cite{1,2,3}. However, the microscopic inhomogeneity of biological tissues severely impedes light focusing inside the tissues: multiple scattering of photons inside the tissues leads to random distortions of the wavefront, which makes it difficult to focus light beyond the optical diffusion limit (1 mm) in biological tissues\cite{1,4,5}.

In recent years, wavefront shaping emerges as a promising method to address this challenge. By modulating the wavefront of the incident light, wavefront shaping compensates for the distortions during the scattering process in a strong scattering environment, thus realizing the refocusing of light at the target position\cite{6,7,8}. Although widely demonstrated for light focusing through diffusers\cite{8,9,10,11}, however, the application of wavefront shaping in tissue \textit{in vivo} has been limited. This is mainly caused by tissue dynamics, where tissue movements such as respiration and blood flow rapidly change the propagation path of light within the tissue. Upon rapid decorrelation of the speckle pattern at millisecond level\cite{12,13,14}, there is a mismatch between the optical transmission characterized using wavefront shaping algorithms and the actual optical transmission, resulting in significant degradation of light focusing performance over time.

Completing wavefront shaping before significant decorrelation has proven to be a promising approach. Numerous studies have demonstrated that millisecond-scale optical focusing through biological tissue can be achieved using digital optical phase conjugation (DOPC)\cite{6,10,11,12,17,19}. In DOPC, a digital camera records the distorted wavefront emerging from the scattering medium, and a spatial light modulator (SLM) projects its phase-conjugated counterpart back to reconstruct the optical focus. With ultrafast, single-shot implementations, DOPC has achieved sub–10 ms runtimes, enabling focusing through millimeter-thick tissues\cite{13,20}. The transmission matrix (TM) method offers an alternative by measuring a set of incident-scattered light field pairs and characterizing the complex light transmission with a linear mapping matrix. Through the calculation of the conjugate transpose of the matrix, the optimal incident wavefront can be achieved to shape the scattered light into arbitrary pattern such as a tightly focused light spot. A series of approaches employing high-speed algorithms \cite{28,29,30,31} and high-speed devices \cite{32,33,34,35,36} have been studied for rapid TM characterization of dynamic media. Compared to DOPC approaches that employ guidestars for light modulation to achieve focusing inside biological tissues \cite{15,23,24}, TM enables retrieval of light-field information from intensity-based signals and can therefore be integrated with absorptive guidestars such as photoacoustics and fluorescence for imaging optimisation \cite{28,boniface2020non,mounaix2018transmission}. 

Traditional TM relies on interferometric approaches to reconstruct the phase information of the output fields, requiring at least N measurement patterns to model a matrix with N input modes\cite{6,13,37,38,39}. When combined with guidestars that provide only light intensity \cite{40,41,42} for non-invasive light focusing, full complex-valued TM reconstruction requires either a phase-shifting technique demanding at least 3N measurements\cite{42,43}, or computational phase retrieval algorithms demanding long processing times\cite{30}. As such, given the same SLM and data processing hardware, reducing N is important for achieving faster TM acquisition and processing for light focusing through dynamic media. In 2021, we simplified the TM characterization process by encoding phase and amplitude information into a real-valued intensity transmission matrix (RVITM), reducing the required measurement patterns to 2N via matrix manipulation\cite{28}. Using a digital micromirror device (DMD) with 4,096 input modes, the RVITM achieved rapid light focusing with a runtime of 300 ms\cite{44} through a static diffuser.

In this work, we investigated the RVITM for light focusing through dynamic media. To accelerate implementation, the RVITM was further simplified by reducing the number of measurement patterns to N and N/2. This approach enables faster acquisition, mitigating decorrelation-induced distortion in dynamic media, albeit at the cost of a reduced static enhancement (defined as the maximum enhancement factor observable in static media). We systematically studied this inherent trade-off to determine optimal settings matched to the temporal stability of different scattering media. Experimental results revealed that achieving a higher static enhancement provides superior performance under slow temporal dynamics, whereas shorter acquisition times are more advantageous in highly dynamic regimes. Using the identified optimal settings, RVITM-based light focusing was further demonstrated on \emph{ex vivo} animal tissue with decorrelation times comparable to those of living tissue, underscoring its potential for future \textit{in vivo} applications.

\begin{figure}[h]
    \centering
    \begin{minipage}{\textwidth}
        \centering
        \includegraphics[width=0.9\textwidth]{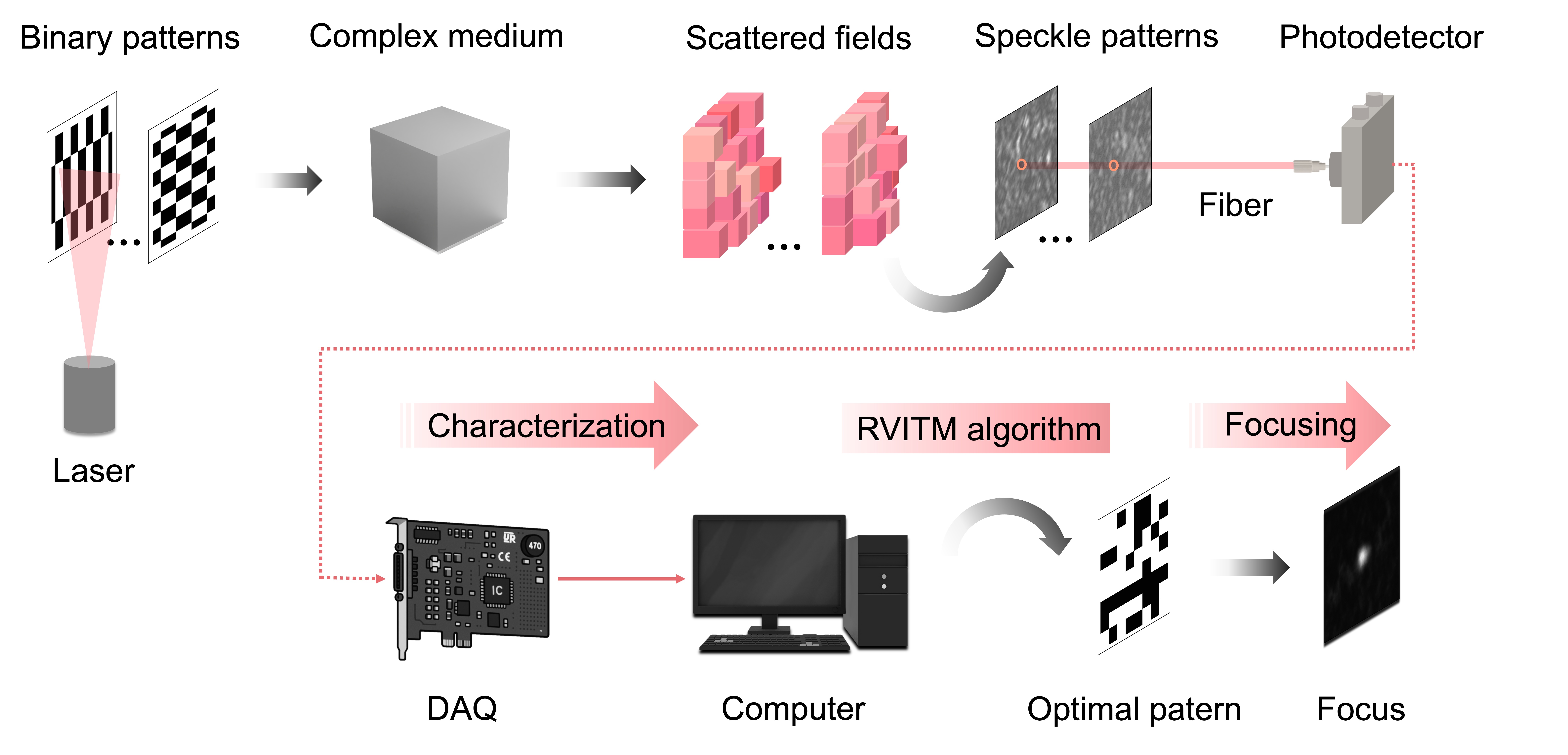}
        \captionsetup{justification=justified, singlelinecheck=false, width=\textwidth}
        \caption{Illustrative diagram of RVITM-based wavefront shaping for light focusing through disordered media. 
        A series of binary Hadamard patterns is used to characterize the scattering medium. 
        After capturing a set of input (DMD patterns) and output (light intensities) pairs, 
        RVITM connected the input light intensities to the output light intensities at the target position (focus). 
        In the focusing step, the optimal DMD pattern was determined by turning `ON` all the DMD micromirrors 
        that cause constructive interference at the target position, thus focusing the light. 
        DAQ: data acquisition card.}
        \label{fig:RVITMdiagram}
    \end{minipage}
\end{figure}

\section{Methods}
\subsection{Principle of RVITM}
The operating scheme of RVITM is depicted in Fig.~\ref{fig:RVITMdiagram}. In the characterization stage, a Hadamard matrix $ \mathrm{H} \in \{-1, +1\} $ with a size of $\mathrm N \times \mathrm N$ was used to generate two binary matrices $ \mathrm{H}_{1} = (\mathrm{H}+1)/2, \mathrm{H}_{2} = (-\mathrm{H}+1)/2 $
. Each column of $[\, \mathrm{H}_{1}, \mathrm{H}_{2} \,]$ was reshaped into a square pattern and displayed on a DMD, whilst the light intensity ($I_n$) at the target position behind the diffuser was recorded. This produced $\mathrm{2N}$ input--output pairs with $\mathrm{N}$ input modes. The light distortion through the diffuser is computationally modeled as:

\begin{equation}
\mathrm{RVITM}_{r} \times [\, \mathrm{H}_{1}, \mathrm{H}_{2} \,] 
= [\, I_{1}, I_{2}, \dots, I_{2\mathrm{N}} \,],
\end{equation}

where $\mathrm{RVITM}_{r}$ represents a row of $\mathrm{RVITM}$. The light intensity achieved with the $k^{\mathrm{th}}$ DMD pattern is expressed as 
$\mathrm{I}_{k} = \left| \sum_{n=1}^{N} \mathrm{t}_{n} \mathrm{E}_{n}^{k} \right|^{2}$, 
where $\mathrm{E}_{n}^{k}$ is either 1 or 0 from $[\mathrm{H}_{1}, \mathrm{H}_{2}]$, representing the $n^{\mathrm{th}}$ input mode in the $k^{{th}}$ DMD pattern; ${t}_{n}$ describes the actual transport distortion, which is usually a complex-valued transmission coefficient in conventional $\mathrm{TM}$. The whole derivation process is detailed in \cite{44,X1}; here briefly, the $\mathrm{RVITM}_{r}$ can be calculated via matrix manipulation and each element $\mathrm{RVITM}_{r}^{n}$ is expressed as:
\begin{align}
\mathrm{RVITM}_r^n = 2 \Bigg[ 
& |t_n|^2 \sum_{k=1}^{2N} (E_n^k)^2 h_n^k  + \sum_{\substack{i=1 \\ i \neq n}}^{N} |t_i|^2 
    \sum_{k=1}^{2N} |E_i^k|^2 h_n^k \notag \\
& + \sum_{\substack{i=1 \\ i \neq n}}^{N} 
    (t_n t_i^* + t_n^* t_i) 
    \sum_{k=1}^{2N} (E_n^k E_i^k) h_n^k \notag \\
& + \sum_{\substack{i=1 \\ i \neq n}}^{N} 
    \sum_{\substack{j=2 \\ j \neq n, j > i}}^{N} 
    (t_i t_j^* + t_i^* t_j) 
    \sum_{k=1}^{2N} (E_i^k E_j^k) h_n^k 
\Bigg],
\label{eq:RVITM}
\end{align}
where $\mathrm{h}_{n}^{k}$ represents the corresponding element in $[\mathrm{H}, -\mathrm{H}]$, defined by $\mathrm{h}_{n}^{k} = 2\mathrm{E}_{n}^{k} - 1$. As detailed in \cite{44,X2}, the fully sampled binary Hadamard matrix $[\mathrm{H}_{1}, \mathrm{H}_{2}]$ allows selective cancellation of interference terms whose $\mathrm{h}_{n}^{k}$ encodings contain equal numbers of $+1$ and $-1$. Consequently, Eq.~\eqref{eq:RVITM} can be reformulated as:


\begin{equation}
\begin{aligned}
\mathrm{RVITM}_{r}^{n} 
&= 2 \Bigg[ N|t_{n}|^{2} 
+ \frac{N}{2} \sum_{\substack{i=1 \\ i \neq n}}^{N} 
\left( t_{n} t_{i}^{*} + t_{n}^{*} t_{i} \right) \Bigg] \\
&= N \mathrm{A}_{R} \mathrm{A}_{n} \cos(\theta_{n} - \varphi_{R}),
\end{aligned}
\label{eq:RVITM3}
\end{equation}

where $\mathrm A_{n}$ and $\mathrm \theta_{n}$ represent the amplitude and phase of $t_{n}$, 
$\mathrm{A}_{R}$ and $\varphi_{R}$ are the amplitude and phase of the reference field $\mathrm{R}$ 
(typically measured when all DMD mirrors are ‘ON’). The optimal DMD pattern is achieved by turning ‘ON’ micromirrors with positive $\mathrm{RVITM}_{r}^{n}$, 
indicating pure constructive interference $(\mathrm\theta_{n} - \varphi_{R} \in [-\pi/2, \pi/2])$ at the target position for light focusing.

The runtime of $\mathrm{RVITM}$ is dependent on the number of input mode $\mathrm N$. While reducing $\mathrm N$ decreases the number of measurement patterns and thus improves the speed of $\mathrm{RVITM}$, 
it also reduces the static enhancement factor, as defined by $\mathrm\eta = N / 2\pi$ \cite{45}. Another strategy for reducing system runtime is decreasing the number of measurements to $\mathrm N$ and $\mathrm N/2$ 
by omitting $\mathrm{H}_{2}$ and removing half of $\mathrm{H}_{1}$. 
Accordingly, the light distortion through the diffuser is computationally modeled as:
\begin{align}
\mathrm{RVITM}_{rN} \times [\mathrm{H}_{1}] 
    &= [I_{1}, I_{2}, \cdots, I_{N}], \label{eq:RVITM4} \\
 \mathrm{RVITM}_{\tfrac{rN}{2}} \times 
    \left[ \mathrm{H}_{\tfrac{1}{2}} \right] 
    &= [I_{1}, I_{2}, \cdots, I_{\tfrac{N}{2}}], \label{eq:RVITM5}
\end{align}
where $\mathrm{RVITM}_{rN}$ and $\mathrm{RVITM}_{\tfrac{rN}{2}}$ represent the characterized $\mathrm{RVITM}_{r}$ with $\mathrm N$ and $\mathrm N/2$ patterns, respectively. As such, certain terms that were originally canceled in Eq.~\eqref{eq:RVITM} due to the symmetric encoding of measurement patterns are now retained, and $\mathrm{RVITM}_{r}^{n}$ achieved in these schemes becomes:
\begin{equation}
\mathrm{RVITM}_{r}^{n} 
= \mathrm{A}_{R} A_{n} \cos(\theta_{n} - \varphi_{R}) + \varepsilon_{n},
\label{eq:RVITM6}
\end{equation}
where $\mathrm{\varepsilon}_{n}$ represents the residual cross-interference terms arising from incomplete sampling of the Hadamard matrix. As a result, characterization errors are encoded into the optimal DMD patterns through $\mathrm{RVITM}_{rN}$ and $\mathrm{RVITM}_{\tfrac{rN}{2}}$, thereby degrading light focusing performance.

\subsection{RVITM characterization through static scattering media}
The fundamental trade-off between wavefront shaping runtime and enhancement was evaluated on a static diffuser. System runtime was defined as the time required to characterize the scattering medium and compute the optimal pattern for DMD upload to achieve optical focusing. The enhancement factor ($\eta$) was calculated as the ratio between the focal intensity obtained with the optimal pattern and the averaged intensity from a series of random patterns with the same number of modes activated. Focusing performance was assessed for different input mode counts (N=256, 1024, and 4096) under three measurement schemes: the full 2N scheme, reduced N, and N/2 schemes. To maintain a constant illumination area (16384 micromirrors), different numbers of DMD micromirrors were grouped into a single macro-pixel, controlled in an “ON” or “OFF” state. The corresponding macro-pixel sizes for N=256, 1024, and 4096 were 64, 16, and 4 micromirrors, respectively. For each configuration, sixteen measurements were taken at different diffuser locations and averaged.

\subsection{Light focusing through dynamic media}
\begin{figure}[h]
    \centering
    \begin{minipage}{\textwidth}
        \centering
        \includegraphics[width=1\textwidth]{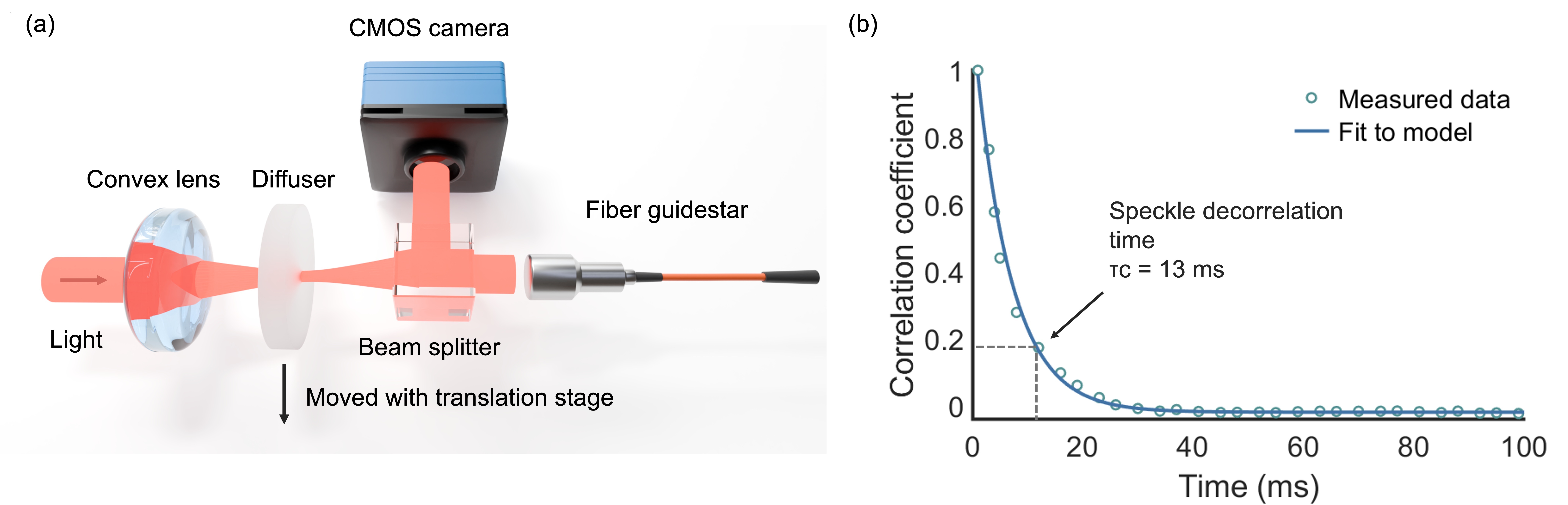}
        \captionsetup{justification=justified, singlelinecheck=false, width=\textwidth}
        \caption{Measurement of speckle decorrelation time in dynamic media. (a) Moving diffuser setup. (b) Correlation coefficient between the speckle patterns over time, when a 2 mm thick diffuser was moved at 0.40 mm/s. Speckle decorrelation time $\tau_{c}$ = 13 ms was determined for this speed.}
        \label{fig:expdec}
    \end{minipage}
\end{figure}

The performance of light focusing through dynamic media was evaluated under controllable dynamic conditions. As illustrated in Fig. \ref{fig:expdec}(a), a glass diffuser was mounted on a motorised XY translation stage (M30XY/M, Thorlabs, New Jersey, USA). By varying the translation speed from 0.01~mm/s to 0.50~mm/s, different levels of tissue dynamics were simulated, corresponding to speckle decorrelation times ranging from 300~ms to 10~ms. A camera recorded sequential speckle patterns at each speed, whilst the temporal correlation coefficient $\mathrm{y}(t)$ between each speckle frame and the initial frame was computed and fitted versus time using $\mathrm{y}(t) = \exp(-2t^{2}/\tau_{c}^{2})$~\cite{46}. The decorrelation time $\tau_{c}$ was calculated by defining the time at which the correlation coefficient decreases to $1/e^{2}$ or 13.5\%. For example, Figure~2b shows the speckle correlation coefficient over time when the diffuser was moved at 0.40~mm/s, from which $\tau_{c} = 13$~ms was determined. Focusing performance was further evaluated on \textit{ex vivo} biological tissues, specifically porcine liver and chicken breast (Fig.~\ref{fig:expdec}(a)). These tissues were sliced into $\sim 2$~mm thickness sections and placed on glass slides, providing fast-decorrelating biological environments.

\subsection{Experimental setup}
\begin{figure}[htbp]
    \centering
    \begin{minipage}{\textwidth}
        \centering
        \includegraphics[width=1\textwidth]{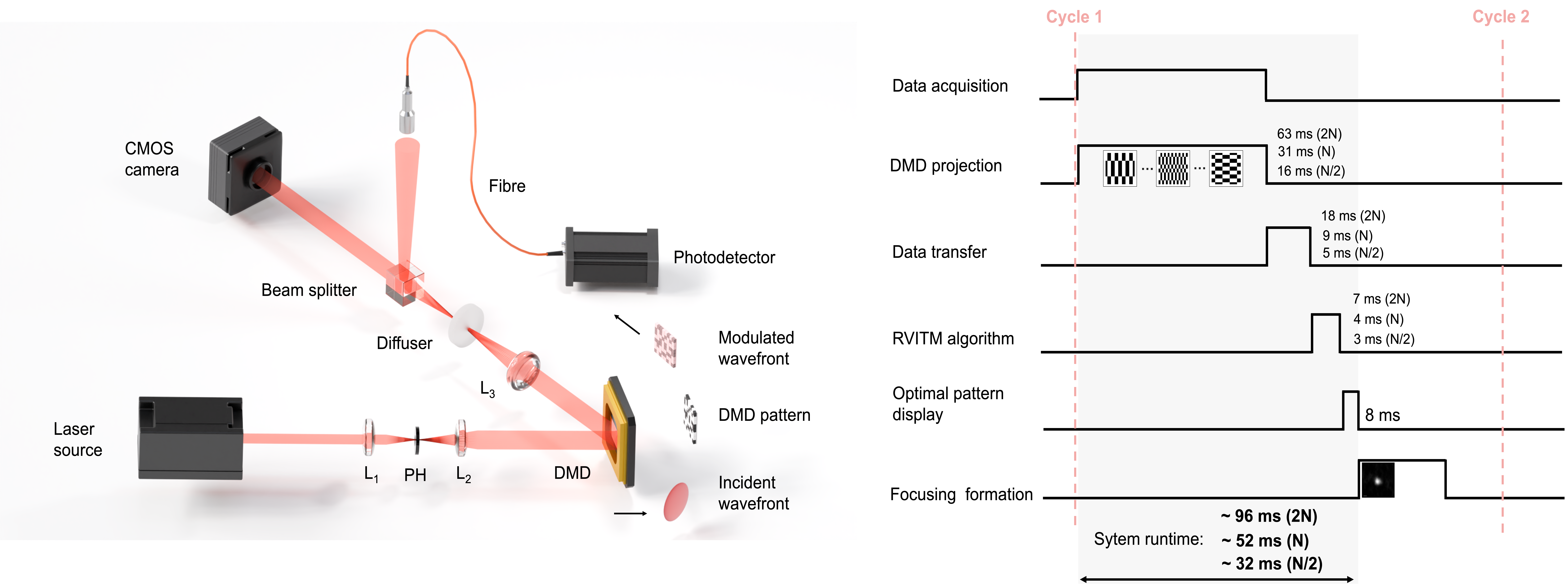}
        \captionsetup{justification=justified, singlelinecheck=false, width=\textwidth}
        \caption{Experimental setup and workflow of RVITM-based wavefront shaping system. (a) Schematic of the experimental setup. L$_{1-3}$, convex lenses; DMD, digital micromirror device; PH, pinhole. (b) Workflow of the high-speed wavefront shaping system using full 2N scheme, reduced N and N/2 scheme for focusing light through scattering medium, N = 1024.}
        \label{fig:expset}
    \end{minipage}
\end{figure}

The experimental setup is illustrated in Fig.~\ref{fig:expset}(a). A pulsed laser source (Onda 1064, BrightSolutions, Prado PV, Italy) emitting at 1064 nm with a pulse width of 5 ns was used as the illumination source. The laser beam was reflected off a digital micromirror device (DMD; DLP650LNIR, 1280 × 800 pixels, Texas Instruments, USA), which functions as a binary spatial light modulator operating at 38 kHz. The modulated beam was then focused by a convex lens (AC254-050-A-ML, Thorlabs, New Jersey, USA) and passes through a ground glass diffuser (N-BK7 Ground Glass Diffuser, 220 Grit, Thorlabs, New Jersey, USA). A beam splitter (BS068, Thorlabs, New Jersey, USA) was positioned behind the diffuser to divide the scattered light into two paths. One path was directed to an avalanche photodetector (APD410A/M, Thorlabs, USA) through an optical fiber (Ø100µm, 0.29 NA, 1 m, F-MLD-C-1FC, Newport, Cheshire, United Kingdom), which measured the light intensity at the target point as a guidestar. The other path was sent to a Complementary Metal Oxide Semiconductor (CMOS) camera (acA1300-200um, BASLER, Germany) for monitoring the wavefront shaping focusing performance. The intensity signals acquired by the photodetector were digitized using a high-speed data acquisition card (DAQ) (M4i.4421, Spectroscopy Instruments, Germany) and transferred to a personal computer for processing. Synchronization between the laser, DMD, and photodetector, was controlled by an arbitrary waveform generator (33600A, Keysight, Santa Rosa, California). 

\section{Results}
\subsection{RVITM characterization}
\begin{figure}[h]
    \centering
    \begin{minipage}{\textwidth}
        \centering
        \includegraphics[width=0.9\textwidth]{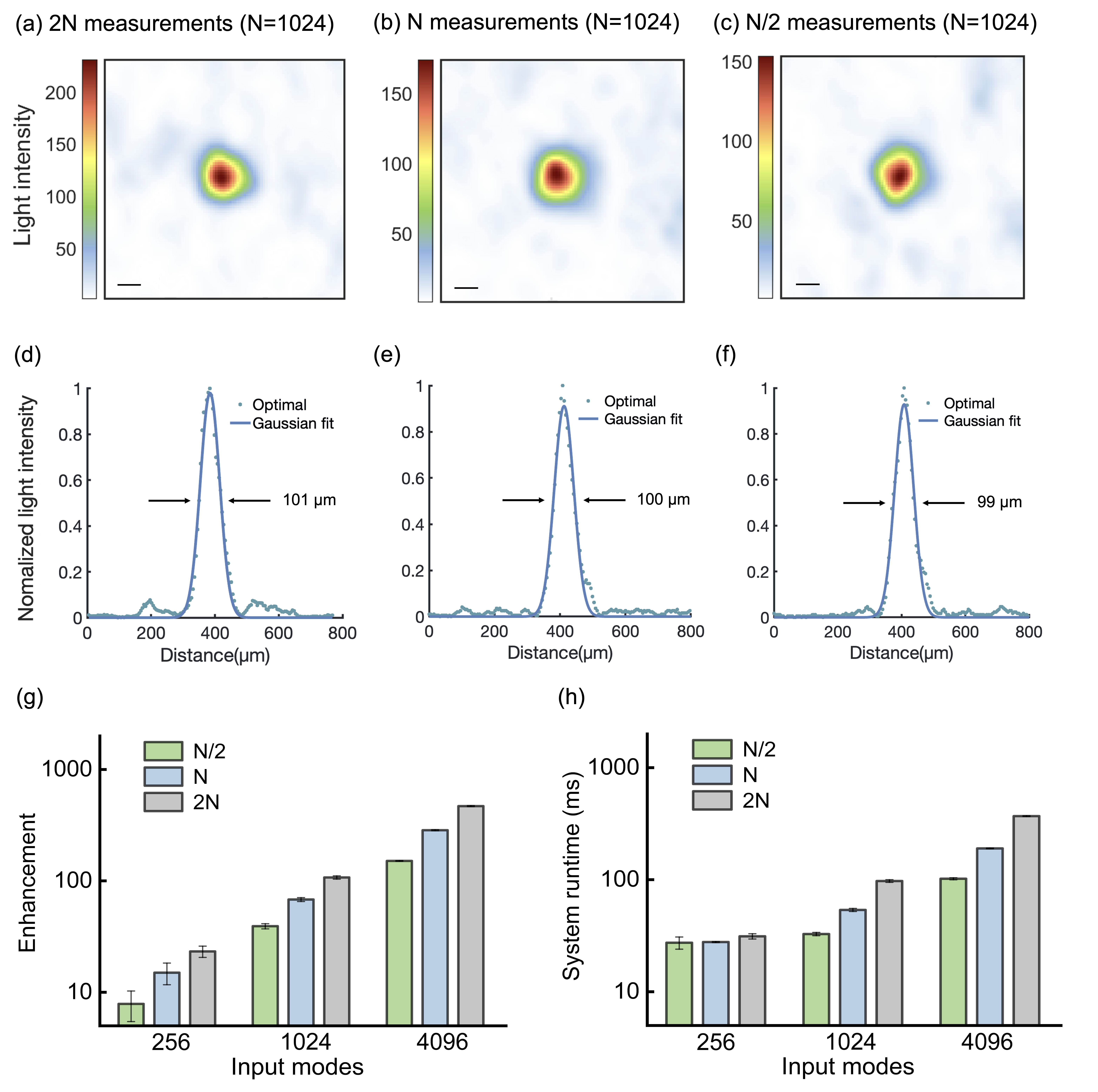}
        \captionsetup{justification=justified, singlelinecheck=false, width=\textwidth}
        \caption{Focusing performance of RVITM through static media. (a) - (c) The focus in the image plane after (a) RVITM modulation (2N measurements), (b) RVITM modulation (N measurements), (c) RVITM modulation (N/2 measurements). (d) - (f) Cross section profile at the largest pixel value of the focus image in (a) - (c). (g) System runtime for different mode and number of measurements settings. (h) Focus enhancement for different mode and number of measurements settings. Scale bar is 50 µm.}
        \label{fig:Sta}
    \end{minipage}
\end{figure}

The light focusing through a static diffuser was achieved using RVITM with different settings. As an example shown in Figs.~\ref{fig:Sta}(a)-\ref{fig:Sta}(c), with N=1024 under the 2N, N and N/2 measurement schemes, random speckle patterns were shaped into tightly focused light spots. The focus geometry, defined by a Gaussian fit to the full width of the half-peak of intensity along horizontal direction, yielded diameters of 100±1 µm (Figs.~\ref{fig:Sta}(d)-\ref{fig:Sta}(f)), consistent with the diameter of the guidestar fiber (100 µm). 

As shown in Fig.~\ref{fig:Sta}(g), changing the number of input modes by a factor of four produced a nearly proportional change in enhancement, consistent with the theoretical enhancement factor. Reducing the number of measurement patterns from 2N to N/2 resulted in an approximately linear decrease in enhancement for static media. The highest enhancement (~600) was achieved with the largest number of input modes (N=4096) and the full 2N measurement scheme, while the shortest system runtime (27 ms) was obtained with the fewest input modes (N=256) and the N/2 scheme. The lower experimental enhancements compared with theoretical predictions were mainly attributed to system instability and fluctuations in laser energies.

The time cost of each step in the RVITM workflow was quantified, with an example for light focusing with N=1024 shown in Fig.~\ref{fig:expset}(b). The characterization stage, involving projection of the measurement patterns, accounted for the largest fraction of the runtime. This duration decreased when the number of measurements was reduced from 2N to N/2 (from 63/96 to 16/32), while the reduction was limited by fixed system overheads such as DMD loading time, DAQ latency, and software communication delays. Both reducing N and switching from the 2N to N and N/2 schemes lowered the total runtime, however, the runtime remained relatively stable at N=256, indicating that fixed overheads were the dominant factor in this case. 

\begin{figure}[h]
    \centering
    \begin{minipage}{\textwidth}
        \centering
        \includegraphics[width=0.9\textwidth]{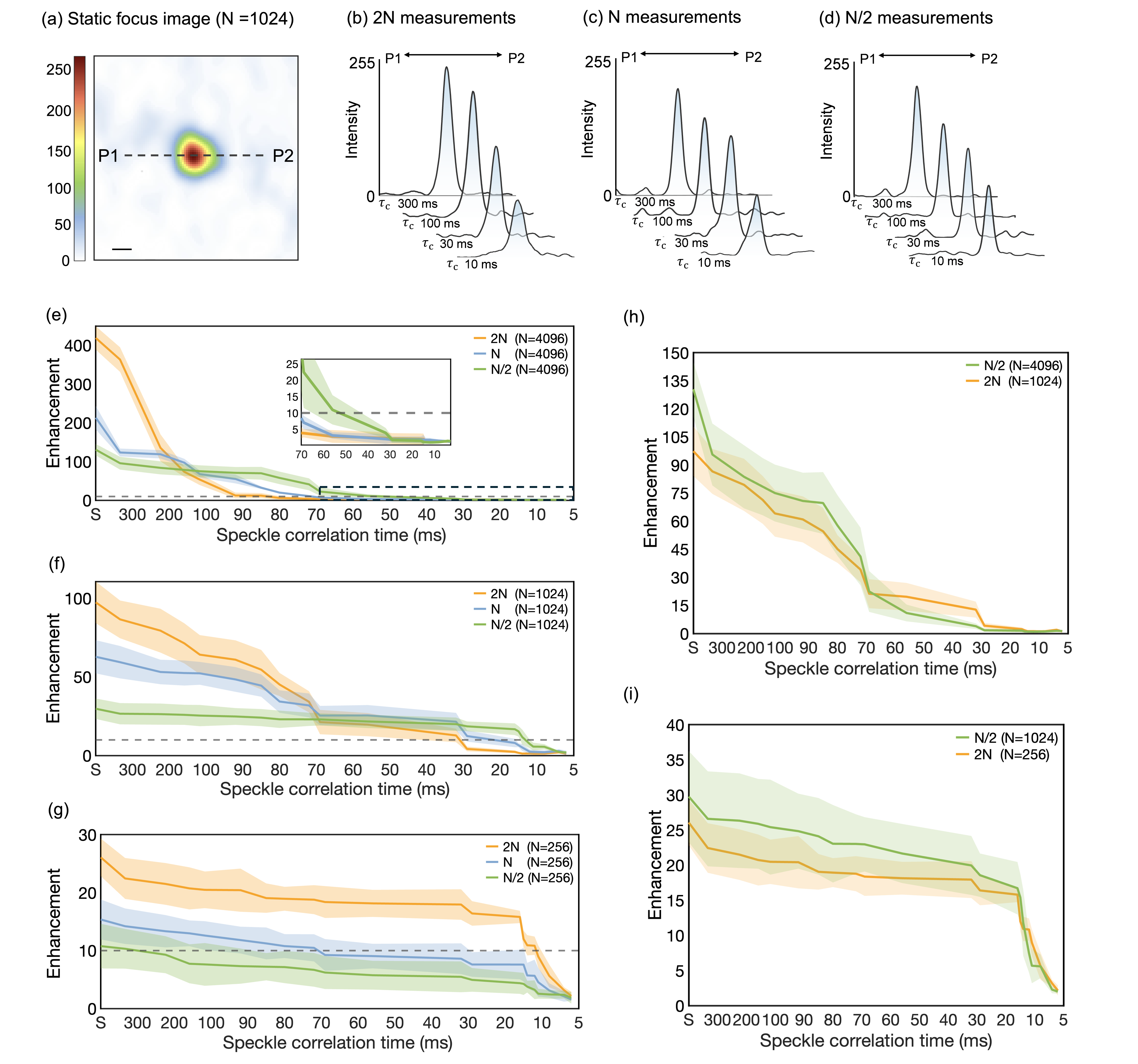}
        \captionsetup{justification=justified, singlelinecheck=false, width=\textwidth}
        \caption{Light focusing performance through dynamic diffusers. (a) The static focus image taken by the CCD camera and the black dash-line shows the region along which 1D profiles are obtained by scanning. Scale bar: 50µm. (b) - (d) The 1D profile across the line from P1 to P2, as shown in panel (a), after continuous RVITM optimization with 2N, N, N/2 measurements when the diffuser is moved on a translational stage to create a series of different correlation times. (e) - (g) Curves of focus enhancement with speckle decorrelation time for different number of measurements with e) 4096 input modes, (f) 1024 input modes and g) 256 input modes. (h) - (i) Curves of focus enhancement with speckle decorrelation time for different number of input modes with same number of measurements: (h) 2048 measurements and (i) 512 measurements.}
        \label{fig:Dy}
    \end{minipage}
\end{figure}

\subsection{Focusing light through moving diffuser}
Light focusing through a moving diffuser using RVITM with 1024 input modes were recorded by the camera, as shown in~\ref{fig:Dy}(a). The focal intensity profiles remained consistence under various speckle decorrelation times, as shown in Figs.~\ref{fig:Dy}(b)-\ref{fig:Dy}(d).  

The speckle decorrelation time of dynamic media showed a significant impact on light focusing enhancement. Within the decorrelation time range from infinity (static) to ~5 ms, light focusing enhancement decreased with increasing media dynamics. These results indicate that schemes with higher static enhancement but slower acquisition are more effective in slowly varying media, whereas faster schemes with lower static enhancement are more suited for rapidly decorrelating environments. Specifically, in the N=4096 case (Fig.~\ref{fig:Dy}(e)), the 2N scheme (369 ms) outperformed the others when the speckle correlation time exceeded 150 ms. Under faster dynamic conditions, however, the N/2 scheme (102 ms) became more effective, maintaining enhancement above10 even at decorrelation times as low as 55 ms. In the N=1024 case (Fig.~\ref{fig:Dy}(f)), the crossover point between schemes shifted to 70 ms due to the increased system speed from fewer modes, with the N/2 scheme (33 ms) sustaining effective focusing down to 12 ms. By contrast, when using only the 256 modes (Fig.~\ref{fig:Dy}(g)), the 2N scheme consistently delivered superior focusing performance across the entire dynamic range, as all schemes have comparable runtime while the 2N scheme has the highest enhancement (~\ref{fig:Sta}(g)).

Notably, RVITM configurations with nearly identical static enhancement and runtime, such as N=4096 with N/2 measurements (runtime 102 ms) and N=1024 with 2N measurements (runtime 97 ms), produced comparable enhancement across a range of speckle correlation times (Fig.~\ref{fig:Dy}(h)). Both settings involved the same number of measurement patterns displayed on DMD. The former yielded slightly higher focusing enhancement and a marginally longer runtime (Figs.~\ref{fig:Sta}(g) and \ref{fig:Sta}(h)), resulting in higher enhancement before the crossover decorrelation time at 70 ms. A similar trend was observed in Fig.~\ref{fig:Dy}(i), where N=1024 with N/2 measurements (33 ms) and N=256 with 2N measurements (31 ms) achieved comparable performance.

\begin{figure}[h]
    \centering
    \begin{minipage}{\textwidth}
        \centering
        \includegraphics[width=0.9\textwidth]{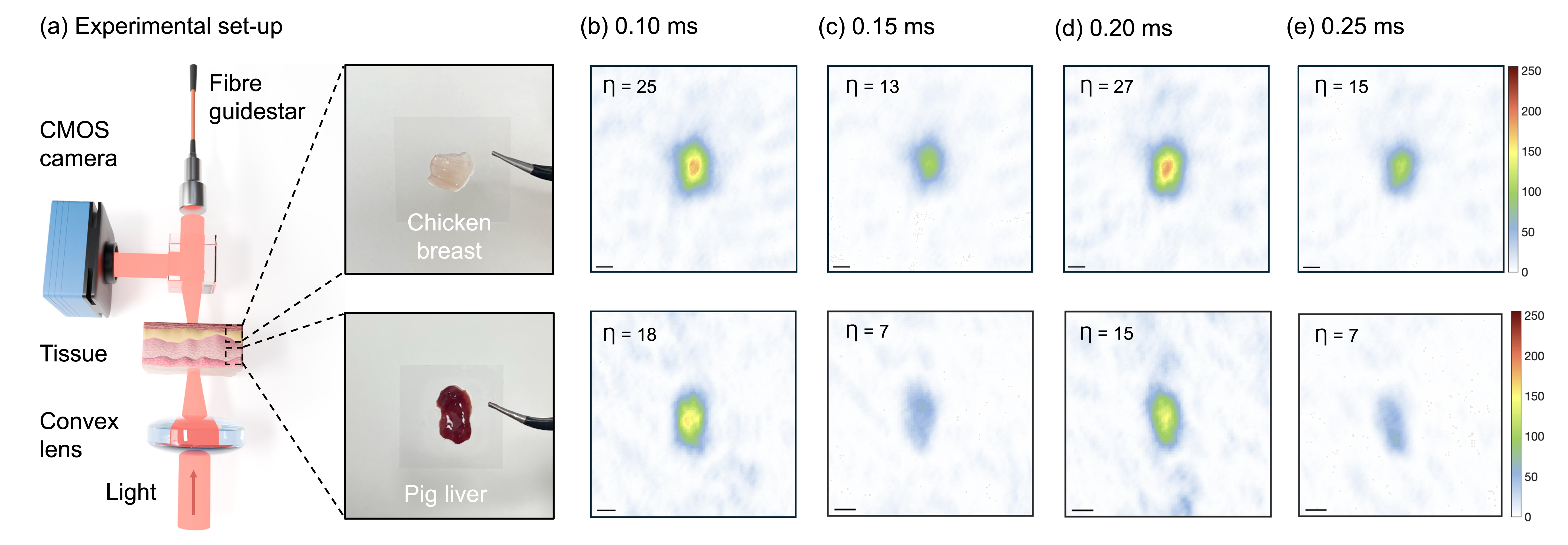}
        \captionsetup{justification=justified, singlelinecheck=false, width=\textwidth}
        \caption{Light focusing performance through biological tissue. (a) Experimental set-up for capturing focus through dynamic tissue using RVITM. (b) - (e) focus images captured from the camera at 0.10 ms, 0.15 ms, 0.20 ms and 0.25 ms using chicken breast (first row) and porcine liver (second row). Scale bar is 50 µm.}
        \label{fig:tissue}
    \end{minipage}
\end{figure}

\subsection{Focusing light through dynamic biological tissue}
The light focusing performance of RVITM was further evaluated on \emph{ex vivo} biological tissues. Tissue samples were attached to a glass coverslip and placed in a vertical orientation, allowing slow gravitational movement. Speckle decorrelation times were measured prior to wavefront shaping, yielding 42 ms for chicken breast and 13 ms for porcine liver. Based on the superior performance observed within this decorrelation range (Figs.~\ref{fig:Dy}(f) and \ref{fig:Dy}(i)), the N/2 scheme with 1,024 input modes was selected. This setting provided a runtime of 33 ms and an enhancement factor of 39 through static diffuser. Representative optical foci are shown in Fig.~\ref{fig:tissue}. Despite the short speckle decorrelation times, optical foci with enhancement factors exceeding 25 and 15 were achieved through chicken breast and porcine liver, respectively. The foci were maintained over time, with the enhancement approximately halving within 0.05 ms after the focusing implementation.

\section{Discussion}
This study demonstrates the RVITM-based wavefront shaping for light focusing through dynamic scattering media across a broad range of decorrelation times. As a TM variant, the full characterization of RVITM reduces the required measurement patterns to 2N, compared with at least 3N in three-phase-shift interferometry \cite{44,46}. The number of measurement patterns can be further reduced to N/2, enabling faster implementations to reduce dynamic perturbations. The fastest RVITM implementation achieved a runtime of 31 ms with an enhancement factor of 12 at a speckle decorrelation time of 10 ms on a moving diffuser, whilst enhancements exceeding 15 were obtained with a 39 ms runtime through porcine liver at a decorrelation time of 13 ms. While not as fast as DOPC, RVITM supports intensity-based feedback such as photoacoustic and fluorescence guidestars. 

The inherent trade-off between static enhancement and implementation runtime was systematically characterized across dynamic media spanning a wide range of decorrelation times. It is indicated that higher static enhancement are more suited to media with slow temporal variations, where the system can exploit the maximum achievable enhancement before decorrelation becomes significant. In contrast, in rapidly decorrelating environments, faster acquisition schemes achieved by reducing the number of measurement patterns outperform higher-enhancement approaches, as they are less affected by temporal distortion despite sacrificing some of the static enhancement factor. These findings highlight the importance of tailoring RVITM configurations to the specific dynamical properties of the scattering medium under study, and  establish a flexible framework for optimizing RVITM in dynamic biological environments, thereby offering practical guidance for efficient wavefront shaping in fast optical focusing scenarios.

While this work demonstrates light focusing through dynamic diffusers using RVITM, further optimization is required for \textit{in vivo} implementation. First, although the RVITM algorithm proved effective in \emph{ex vivo} biological samples with decorrelation times comparable to those of live tissue, \emph{in vivo} environments may present more complex and heterogeneous scattering dynamics. This warrants experimental validation of the RVITM-based acceleration strategy in live animal models and human tissue. Second, implementation speed could be further improved by integrating RVITM with faster spatial light modulators, such as MEMS-based SLMs or 1D grating light valves \cite{36}, and hardware acceleration platforms like field-programmable gate arrays (FPGAs). Third, this study employed a fiber-based guidestar to provide stable and controllable intensity-only feedback. Given that RVITM relies solely on intensity-based feedback, it is inherently compatible with non-invasive photoacoustic guidestars. Building on our previous work, combining RVITM with photoacoustic-guided wavefront shaping (PAGS) could enable non-invasive, deep-tissue imaging in dynamic biological environments. Together, these advancements would move the field closer to real-time, non-invasive optical control in complex \emph{in vivo} settings, with significant implications for translational biomedical imaging and therapeutic applications.

\section{Conclusion}
In conclusion, this study demonstrated light focusing through dynamic media using the RVITM approach. A flexible strategy was introduced to guide optimization of RVITM setting across varying tissue dynamics. These findings provide practical guidance for implementing real-time wavefront shaping in complex biological environments and establish a strong foundation for developing fast, adaptive wavefront shaping systems for clinical and biomedical applications.

\begin{backmatter}
\bmsection{Funding}
Royal Academy of Engineering under Research Fellowship program (RF2223-22-170) and EPSRC Research Council (EPSRC DTP: EP/W524475/1).


\bmsection{Disclosures}
The authors declare no conflicts of interest.

\bmsection{Data availability} 
The data that support the findings of this study are available from the corresponding author upon reasonable request.

\end{backmatter}

\bibliography{Refrences}

\end{document}